\documentclass[aps,prl,twocolumn,preprintnumbers,superscriptaddress,floatfix]{revtex4}

\usepackage{amssymb,multirow}
\usepackage{graphicx}

\begin{document}

\newcommand{\Tr}{\mbox{Tr\,}}
\newcommand{\beq}{\begin{equation}}
\newcommand{\eeq}{\end{equation}}
\newcommand{\bea}{\begin{eqnarray}}
\newcommand{\eea}{\end{eqnarray}}
\renewcommand{\Re}{\mbox{Re}\,}
\renewcommand{\Im}{\mbox{Im}\,}

\voffset 1cm

\newcommand\sect[1]{\emph{#1}---}

\pagestyle{empty}

\preprint{
\begin{minipage}[t]{3in}
\begin{flushright} NI-07-062
\\
SHEP-07-40
\\[30pt]
\hphantom{.}
\end{flushright}
\end{minipage}
}

\title{A Holographic Model Of Hadronization}

\author{Nick Evans and Andrew Tedder}

\affiliation{School of Physics and Astronomy, University of
Southampton,
Southampton, SO17 1BJ, UK}

\begin{abstract}

 We study hadronization of the final state in a
particle-antiparticle annihilation using a holographic gravity
dual description of QCD. At the point of hadronization we match
the events to a simple (Gaussian) energy distribution in the five
dimensional theory. The final state multiplicities are then
modelled by calculating the overlap between the Gaussian and a set
of functions in the fifth dimension which represent each hadron.
We compare our results to those measured in $e^+e^-$ collisions at
LEP and PEP-PETRA. Hadron production numbers, which differ in
range by four orders of magnitude, are reproduced to well within a
factor of two.
\end{abstract}
\maketitle

\sect{Introduction}%
Since  the discovery of the AdS/CFT Correspondence
\cite{malda}, holographic gravitational theories have
been studied to shed light on strongly coupled gauge theories. Phenomenological five dimensional (5D) models in this spirit
(AdS/QCD) \cite{son1} also provide a quantitative description of the QCD meson
spectrum which seem to work at the level of 10\% accuracy. In this
paper we will apply these tools towards another notoriously
difficult QCD calculation: hadronization.

Current Monte Carlo event generator models \cite{PYTHIA} of hadronization 
are complex with
many parameters which are tuned to data in some energy
regime and have limited predictive power. A simpler understanding
of the process would be a boon. Recently, progress has been made
by assuming that after the quarks freeze into hadrons they may be
described as a hadron gas in thermodynamical equilibrium
\cite{thermo}. Such models provide a surprisingly good
description of the multiplicities of hadrons in jets across
several orders of magnitude.

Models of hadronization generically have two parts:
predicting the initial yield of hadrons directly after
annihilation and then allowing for decays of those
particles in transit. Our model, like the thermal model \cite{thermo},
only addresses the first part. Modelling the decays would
involve a theory of branching ratios which we do not
propose. Instead we model this using the available
branching ratio data from collider experiments. 

\begin{figure*}
\includegraphics{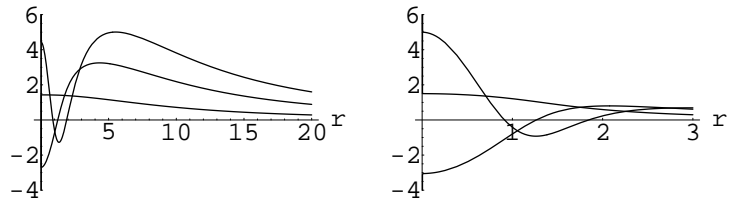}
\caption{\label{functions}.The normalised holographic hadron basis functions for the
$\pi^0$  stack (left) and the $\rho$ meson stack (right)}
\end{figure*}

AdS/QCD models provide a weakly coupled 5D
gravitational theory that describes the hadrons made of confined
quarks. Each hadron and its excited states (e.g. the ``stack" of
the $\rho$ mesons: $\rho(770),\rho(1450),\rho(1700)\ldots$) is
described by a 5D field that shares the Lorentz
properties and global symmetries of the hadron. In the
gravitational theory one seeks solutions for these fields that
separate the 3+1 dimensional dependence from the extra radial
direction dependence, so they take the form $g_n(r) e^{ik_nx}$
with $k_n^2=M_n^2$. There are only regular solutions on the space
for discrete masses corresponding to the meson masses.
The functions $g_n(r)$ form an orthogonal basis.

The dual necessarily only describes the regime of QCD where the
gauge coupling is strong. When asymptotic freedom sets in at short
distances and QCD is best described by free quarks and gluons the
gravitational theory is expected to become strongly coupled itself
and computation becomes impossible. Our description therefore is
dependent on the distribution of energy at the matching energy
scale where hadronization occurs. We will see that a very simple
model reproduces the data very well.

\sect{Hadronization Holographically}%
Our holographic model of hadronization will share with the thermal
models the idea that at the point where the quarks form hadrons,
the energy of the event is democratically available to all
hadronic channels. We describe the initial condition as some
deposition of energy into the 5D model's
stress-energy tensor. The radial dependence of the stress-energy
dependence can be expanded in terms of the functions $g_n$ and
will determine the relative multiplicities of each particle in a
hadron stack. The $x$ dependence will determine the energy and
momentum of the hadrons. In this paper we will simply concentrate
on the multiplicities. There should be some kinematic limit on the
maximum mass of a state produced in the shower
and in
principle this could vary with the centre of mass energy of the event. In practice we shall simply include all known states with mass below 1.7 GeV (but excluding the $a_0(980)$ triplet - it is thought to be a bound kaon state). Above this value the experimental
data on the full spectrum becomes patchy. In addition, the high
mass states in this range are only produced with very small
multiplicities and have a minimal effect on the lighter particle
results.

The result for the multiplicities depends on the choice of the
function expanded in terms of the $g_n$ and this represents the
matching to the underlying asymptotically free QCD dynamics. The
simple guess we will employ is that we should treat all hadronic
channels equally and pick a Gaussian for the initial condition. The height of
the Gaussian determines the absolute value of each particle's
multiplicity and hence is a free parameter which we fit (this parameter can be re-expressed as
the average energy per hadron, $\kappa$, which we detail below). The 
width is also a free parameter which we fit. Finally
the thermal model requires a suppression factor on the production of
strange quarks. We too include a
strangeness suppression factor $\gamma_s$ which multiplies the
Gaussian for each strange quark in the hadron (note that if mixing
occurs the strange quark content need not be an integer).
$\gamma_s$ is another fit parameter in our analysis although 
we find it lies close to one.

The final ingredient we require is a specific AdS/QCD model of the
QCD hadrons. We will adapt a string theory derived model of chiral
symmetry breaking that includes the vector mesons and pions
\cite{shock:CSB}.  That both the vector mesons and the pions are
included is an important feature: the mass spectra of the
pseudo-Goldstone bosons are significantly different to all other
hadrons in QCD. We will then assume that the $g_n$ functions
associated with each hadron stack are not that different from the
$\rho$-stack functions - we will simply reproduce them but with
the mass of the lightest stack member tuned to the experimental
value. Similarly the pion stack can be used to reproduce the
towers of states associated with each pseudo-Goldstone of the
chiral symmetry breaking (i.e. the pions, kaons and eta meson). The relative weighting is parameterised by $R$, a measure of $g_{\mathrm{YM}}^2N$, which we fit.

Putting it all together, we compare our results to that detected at LEP ($\sqrt{s}=91.2$GeV) and PEP-PETRA ($\sqrt{s}= 29$GeV), and find that we can reproduce yields that vary over four orders of magnitude to within a factor of two. It is expected that the biggest source of error comes from the choice of holographic dual.

\sect{Holographic Hadron Basis Functions}
Our holographic model of the hadron spectrum is based on a string
theory construction of a QCD-like gauge theory consisting of a
deformed D3 brane geometry with quarks included through probe D7
branes - the precise details do not need to be understood to follow this letter, but computational details are in the appendix and a more
detailed analysis can be found in \cite{shock:CSB}. The model
realizes dynamical chiral symmetry breaking. The mass of the
lowest lying $\rho$ meson can be dialed by choosing the conformal
symmetry breaking scale in the model, and the mass of the lowest
lying pion can be dialed by choosing the asymptotic quark mass.
The excited states in both stacks are then predicted:
$m_{\rho^*}=1737$ MeV, $m_{\pi^*}=1701$ MeV (c.f. experimental
values of 1459 and 1300 MeV respectively). So whilst they don't
precisely reproduce the experimental values the pattern is at
least roughly right.

As stated above, we assume that the $g_n$
functions associated with each hadron stack are not that different
from the $\rho$-stack functions, and simply rescale  the $r$
coordinate such that the mass of the lowest member of each stack
is correct.

The functions representing the pions, henceforth denoted as $f_n$, are also rescaled. This time we dial the asymptotic quark mass such that the
mass of the lowest member of each stack is correct. We plot the
first three functions for $f_n,g_n$ in figure \ref{functions}.

The derivation of these functions 
is briefly reviewed in the appendix. We stress that here we
seek to describe hadronization not to provide a complete
holographic model of QCD. Hopefully the functions we use,
based on the string model, are a reasonable
phenomenological basis.

\sect{Overlap Computation of Multiplicities}
With our holographic functions $f_n,g_n$ in place for each hadron
stack we can now proceed with computing the expected initial yield
in a hadronisation event.

We assume that all the five dimensional fields $\Psi(r)$ (e.g. each
component of a gauge field $A_{\mu}$ describing the $\rho$ mesons) have
a common initial condition of a Gaussian centred at $r=0$ and of
width $\Lambda$. To find the multiplicities of each
stack member we compute \beq c_n = \int_0^\infty \Psi(r) \:w(r)g_n(r) dr \eeq where $w(r)$
is the weighting function associated with the basis functions
$g_n$. The Gaussian receives most support from the lowest mass states, whilst very highly
excited states have no overlap with the Gaussian because of their 
highly oscillatory nature in the IR. 

The multiplicity is simply given by $c_n^2$ multiplied by (2J+1)
where $J$ is the spin of the hadron.

There are a number of special cases, which we address in turn.

\sect{Pseudo-Goldstone Bosons}
The pseudo-Goldstones are described by a separate holographic
field,  $\theta$, in our model and are represented by functions
$f_n(r)$. The relative contributions the fields make to the
stress-energy tensor are \beq T_{rr} \sim \Delta_1(r) 
(\partial_r \theta)^2+ \Delta_2(r)\eta^{\mu\nu}
(\partial_rA_{\mu})(\partial_rA_{\nu}) + \dots \label{stress-en}\eeq 
$\Delta_1, \Delta_2$ are functions of $r$, calculable from the
holographic model. So that these fields see the same contribution
to the stress energy tensor we therefore rescale the Gaussian. If
our standard Gaussian is $\Psi(r)$ then the pion field sees  $\int
dr \sqrt{\frac{\Delta_2}{\Delta_1}}\partial_r\Psi$.

\sect{Strangeness Suppression Factor}%
Since the underlying asymptotically free dynamics may distinguish
the strange quark from the up and down quarks, we also multiply
the Gaussian by a factor of $(\gamma_s)^\sigma$ where $\sigma$ is
the strangeness content of the stack. $\gamma_s$ is then the
second fit parameter in our procedure. Note that this procedure is
rather crude because different members of a stack may mix to
varying degrees with other states. For example the $\eta(548)$ has
32\% strangeness content while the $\eta^{**}$ has 100\%. In these
cases we set $\gamma_s$ by the strangeness content of the lightest
member of the stack.

\sect{Height of Gaussian}%
The normalisation of the Gaussian tells us the relative
multiplicities of the various hadrons in an event. An overall
multiplicative factor $\kappa$ sets the absolute number of each
species and we fit this value. $\kappa$ determines the total
number of final state particles (before allowing for decays in
transit to the detector), and hence we express it as the average
hadron energy in the collision.

\sect{A Fourth Parameter}
Our choice of holographic dual also contains a free parameter, $R$, 
which sets the 't Hooft coupling in the gravity dual. We fit it to the 
data. However, $R$ is not in the same class as $\Lambda,\gamma_s,\kappa$. 
$R$ has a sound theoretical background, and would be predicted if the holographic dual to QCD was known.

\sect{Decay in transit}
Once we have calculated the initial yield of hadrons, we then have
to allow for decays of the particles in transit from the
interaction point to the detector. Branching ratios are taken from \cite{PDG06},
 and particles that can be detected at LEP (whose results
we will compare to) are set as stable. All the other particles are
allowed to decay through the decay channels until they reach one
of the stable particles. In this way we get a list of numbers
which is what our model predicts would be seen at LEP.

\sect{Predictions}
We compare our results both to $e^+e^-$  collisions performed at
LEP ($\sqrt{s}=$ 91.2 GeV), and at PEP-PETRA ($\sqrt{s}= 29$ GeV). Average
multiplicities of various hadrons have been compiled in
\cite{thermo}, and we reproduce them here, along with our results,
in table \ref{resulttab}. For both sets of results we have performed a four parameter fit 
so as to minimise the rms error. The pion predictions are expected to be low because our 
holographic model predicts over-massive excited states. Thus yields of particles such as 
$\pi(1300)$ and $\rho(1450)$ are unnaturally suppressed which would otherwise be expected 
to give significant contributions to the pion multiplicities. Despite this, the fits are 
very good, with rms errors of 34\% and 46\% for  $\sqrt{s}=$ 91.2 GeV and $\sqrt{s}=$ 29 
GeV respectively. The $\eta'$ yields aren't ideal, but previous models have 
had the same problem \cite{thermo}. Furthermore, our 
model is expected to have poor predictive power for the $\eta'$ meson: technically it is a pseudo-Goldstone boson, but instanton effects cancel out this effect. Our 
model just treats it as non-Goldstone boson, which is probably over simplistic. In 
addition, it mixes heavily with $\eta(548)$: our model contains no good parameterization 
of mixing.

The extra inaccuracy for the PEP-PETRA matching comes, in large part, from the $\Omega$ yield. This has a very large experimental error (see \cite{thermo}), and so we may not be matching to the correct value: on comparing the PEP-PETRA and LEP experimental yields, a decrease by a factor of 10 adds doubt to the accuracy of the PEP-PETRA measurement. 

\begin{table}
\caption{\label{resulttab} Results of the model for hadron yields at $\sqrt{s}=91.2$ GeV (centre column) and $\sqrt{s}= 29$ GeV (right column). The relevant 4 free parameter values are shown in the final row.}
\begin{ruledtabular}
\begin{tabular}{c|cc|cc}
Hadron & Model & Expt & Model & Expt\\\hline
 $\pi ^+$ & 5.95 & 8.5 &4.07& 5.35\\
 $\pi ^0 $& 6.43 & 9.2 &4.41&5.3\\
 $K^+$ & 1.09 & 1.2 &0.68&0.7\\
 $K^0$ &1.09  & 1.0 &0.68&0.69\\
 $\eta$  & 1.06 & 0.93&0.66&0.584 \\
 $\rho ^0$ & 1.33 & 1.2 &0.88&0.9\\
$ K^{*+}$ & 0.387 & 0.36 &0.28&0.31\\
 $K^{*0}$ & 0.385 & 0.37 &0.28&0.28\\
 $\eta' $ & 0.042 & 0.13 &0.03&0.26\\
 p & 0.41 & 0.406 &0.30&0.3\\
 $\phi$  & 0.03 & 0.01 &0.02&0.084\\
 $\Lambda$  & 0.172 & 0.19 &0.13&0.0983\\
 $\frac{\Sigma ^{*+}+\Sigma ^{*-}}{2}$ & 0.0120 & 0.0094 &0.0089&0.0083\\
$ \Xi^-$ & 0.012 & 0.012 &0.0088&0.0083\\
$ \Xi^{*0}$ & 0.0040 & 0.0033&n/a& n/a\\
 $\Omega$  & 0.0011 & 0.0014&0.0008&0.007\\
\hline
$\Lambda \mathrm{\:(MeV)},\kappa \mathrm{\:(GeV)}$& 
\multicolumn{2}{c|}{150, 4.96}& \multicolumn{2}{c}{152, 2.35}\\
$\gamma_s,R$& 
\multicolumn{2}{c|}{0.97, 2.6}& \multicolumn{2}{c}{0.97, 2.4}\\
\end{tabular}
\end{ruledtabular}

\end{table}

\sect{Conclusions}
We assumed that every hadron in QCD can in principal be
represented by a function in the $r$ coordinate of the 5D
holographic theory of QCD. We then proposed that hadronization can
be modelled by hypothesising that the initial yield (that is
before the particle created starts decaying) for any hadron is
given by the square of the overlap between the function which
represents the hadron, and a Gaussian, centred at the origin, with
a width of $\Lambda$. In addition we have two other
parameters in the theory; a strangeness suppression factor to
account for the heaviness of the strange quark, and $\kappa$ which
determines with what energy the particles leave the interaction
point.

With the full holographic dual to QCD currently unknown, we made
some reasonable assumptions to achieve a full set of functions which
represent every hadron. We then compared the results to $e^+e^-$ 
collisions made at LEP
and PEP-PETRA. The results are surprisingly good suggesting the broad framework
is correct.

The model of hadronization presented here is applicable to all
particle-antiparticle annihilation events, where the fireball
after the collision has no residual quantum numbers. Broadening
this model to include events such as deep inelastic proton-proton
scattering, and heavy ion collisions would clearly be desirable. A
natural way to do this would be to include enhancement factors on
the multiplicities of stacks contributing to the quantum numbers
that are non zero in the final state. We leave such an analysis
for the future.

\textit{We would like to thank James Ettle and Francesco Becattini for
their help in performing the branching ratio part of the
calculation, and Ed Threlfall for his help in the preliminary stages of this paper.}

\vspace{-0.5cm}
\section{Appendix}
\vspace{-0.2cm}

\sect{String Theory Progenitor}
The phenomenological model used here is based on the AdS/CFT
Correspondence realization of chiral symmetry breaking in
\cite{shock:CSB}. That model consists of a dilaton flow
deformed AdS geometry
\vspace{-0.3cm}
\begin{displaymath} ds^2 = H^{-1/2} f^{\delta/4} dx_{4}^2 + R^2 H^{1/2} f^{(2-\delta)/4} \frac{w^4 - b^4}{ w^4 } \sum_{i=1}^6 dw_i^2  \end{displaymath}
\vspace{-0.6cm}
\beq \mathrm{where\;\;}  H =  f^{\delta} - 1,\hspace{0.1cm} f=\frac{w^4+b^4}{w^4-b^4}, \hspace{0.1cm} e^{2 \phi} = e^{2 \phi_0} f^{\Delta}\eeq 
\vspace{-0.3cm}

There are formally two free
parameters, $R$ and $b$, since $\delta = 1/ 2 b^4$ and $ \Delta^2 = 10 - \delta^2$. The parameter $b$ sets
the conformal symmetry breaking scale and we set it equal to one from this point onwards. We will use the $\rho(770)$ mass to set an absolute mass scale. $R$ sets the 't Hooft coupling which we treat as a free parameter.

Quarks are introduced by including probe D7 branes into the
geometry. We minimize the D7's world-volume in the spacetime
around the D3 branes. This is encoded by the Dirac Born Infeld
action in the Einstein frame of the D7 brane, which also contains a superpartner U(1) gauge field which
describes the vector mesons. The full action is 
\vspace{-0.2cm}
\beq
  S\sim\int d^8\xi~e^\phi\left[-\det({\bf P}[g_{ab}]+2\pi\alpha'
  e^{-\phi/2}F_{ab})\right]^{\frac{1}{2}} \label{fullact} 
\eeq 
which, expanded to second order gives 
\vspace{-0.2cm}
\bea
  S&&\sim\int d^8\xi~e^\phi\sqrt{-g}(1+\dot\sigma^2)^{\frac{1}{2}}\times \nonumber\\
  &&\left[1+\frac{1}{2} g_{rr}{ \sigma^2 \over (1+\dot\sigma^2)}\partial^a\theta
  \partial_a\theta - \frac{1}{4}{(2\pi\alpha')^2\over (1+\dot\sigma^2)}
  e^{-\phi}F^2\right]. \label{expact}\eea

Our action is defined in 8D, rather than the usual 5D. But since we assume that the fields have 
no components in the 3-sphere (which is appropriate for duals to non-supersymmetric fields), 
(\ref{expact}) can easily be recast as a 5D action.

Substituting from the geometry above we can find the equation of
motion for the radial separation, $\sigma$, of the two branes in
the $8-9$ directions as a function of the radial coordinate $r$ in
the $4-7$ directions. We first calculate the background solution as in \cite{shock:CSB}. The large $r$ asymptotic solutions take the form $\sigma_0 = m +
c/r^2$ with $m$ representing the quark mass and $c$ the quark
condensate. The regular solutions have non-zero $c$ even when
$m=0$ and describe chiral symmetry breaking.

Fluctuations of the brane about $\sigma_0$ in the
angular direction, $\theta$, of the $8-9$ plane correspond to the pion. If we write $ \theta(r,x)=2\pi \alpha'f(r) \sin (kx)$, the equation of motion for $f$ is given by
\vspace{-0.2cm}
\beq \frac{\partial}{\partial r} \left(\frac{e^{\phi} \mathcal{G}}{\sqrt{1+(\dot{\sigma}_0)^2}}\sigma_0^2 \dot{f} \right) + M^2 w_f f =0 \label{pions}
\eeq
\vspace{-0.5cm}
 \beq \mathrm{with\;\;\;}w_f = \frac{R^2 e^{\phi}\mathcal{G}}{\sqrt{1+\dot{\sigma}_0^2}}H f^{1/4}\frac{(r^2+\sigma_0^2)^2-1}{(r^2+\sigma_0^2)^2} \sigma_0^2 \eeq

\vspace{-0.1cm}
Finding regular solutions to (\ref{pions}) is simply a Sturm-Liouville eigenvalue problem, and hence we know that the regular solutions $f_n$ will form a basis under the weighting function $w_f$.

Similarly, if we write $A_{\mu}=g(r) \sin(kx) \epsilon_{\mu}$, the equation of motion for $g$ is 
\vspace{-0.2cm}
\beq \frac{\partial}{\partial r} \left( \frac{e^{\phi}\mathcal{G}}{\sqrt{1+\dot{\sigma}_0^2}}\frac{w^4}{\sqrt{(w^4+1)(w^4-1)}}\dot{g}\right) +  M^2 w_g g =0
\eeq
\vspace{-0.5cm}
\beq w_g = R^2 e^{\phi} \mathcal{G} \sqrt{1+\dot{\sigma}_0^2}H f^{-1/4} \eeq

Where once again $g_n$ form a basis under the weighting function $w_g$.

\sect{Stress-energy tensor}
We also need to know the stress-energy contributions.
\vspace{-0.3cm}
\beq T_{rr} = -\frac{2}{\sqrt{-g}}\frac{\delta}{\delta g^{rr}} \left(e^{\phi}\sqrt{-g} \mathcal{L} \right) \eeq

We have to be careful, and use (\ref{fullact}), not (\ref{expact}), before expanding to second order. On doing so, we find
\vspace{-0.2cm}
\beq  \frac{\Delta_2}{\Delta_1}=\frac{g^{\mu\mu} g^{rr}}{\sigma^2 e^{\phi}} \eeq
with $\Delta_1,\Delta_2$ defined in equation (\ref{stress-en}).

\end{document}